\documentclass[aip,amsmath,amsfonts,amssymb,reprint,floatfix ]{revtex4-1}
\usepackage[a4paper, total={17cm, 23.4cm}]{geometry}
\usepackage{graphicx}
\usepackage{dcolumn}
\usepackage{subcaption}
\usepackage{booktabs}
\usepackage{bm}%
\usepackage[colorlinks=true,linkcolor=blue]{hyperref}%
\usepackage{floatrow}
\usepackage{mathrsfs}
\usepackage{epstopdf}

\newcommand{\SiO}{$\rm{SiO_2}$}
\newcommand{\SiN}{$\rm{SiN_x}$}
\newfloatcommand{capbtabbox}{table}[][\FBwidth]

\expandafter\ifx\csname package@font\endcsname\relax\else
 \expandafter\expandafter
 \expandafter\usepackage
 \expandafter\expandafter
 \expandafter{\csname package@font\endcsname}%
\fi

\begin{document}
\title{First Characterization of a Superconducting Filter-bank
Spectrometer for  Hyper-spectral Microwave Atmospheric Sounding
with Transition Edge Sensor Readout.}
\author{D. J. Goldie}
\email{goldie@mrao.cam.ac.uk.}
\affiliation{
Quantum Sensors Group, Cavendish Laboratory, University of Cambridge, JJ Thomson Av., Cambridge, UK. }  %
\author{C. N. Thomas}
\affiliation{
Quantum Sensors Group, Cavendish Laboratory, University of Cambridge, JJ Thomson Av., Cambridge, UK. }
\author{S. Withington}
\affiliation{
Quantum Sensors Group, Cavendish Laboratory, University of Cambridge, JJ Thomson Av., Cambridge, UK. }  %
\author{A. Orlando}
\affiliation{
Astronomy Instrumentation Group, School of Physics and Astronomy, Cardiff University, Cardiff, UK.}
\author{R. Sudiwala}
\affiliation{
Astronomy Instrumentation Group, School of Physics and Astronomy, Cardiff University, Cardiff, UK.}
\author{P. Hargrave}
\affiliation{
Astronomy Instrumentation Group, School of Physics and Astronomy, Cardiff University, Cardiff, UK.}
\author{P. K. Dongre}
\affiliation{
Astronomy Instrumentation Group, School of Physics and Astronomy, Cardiff University, Cardiff, UK.}
\date{\today}%

\begin{abstract}
We describe the design, fabrication, integration and characterization of a prototype superconducting filter bank with transition
edge sensor readout designed to explore millimetre-wave detection  at frequencies in the range
 40 to $65\,\,\textrm{GHz}$.
Results indicate highly uniform filter channel placement in frequency and high overall detection efficiency. The route to a
full atmospheric sounding instrument in this frequency range is discussed.
\end{abstract}
\maketitle

\section{Introduction}
Superconducting on-chip filter-bank spectrometers (SFBSs)  are a promising technology
for a number of  scientifically important
applications in astronomy and meteorology that require low-noise,
spectroscopic measurements at millimetre and sub-millimetre wavelengths.
SFBSs are capable of achieving high channel counts and
the individual channel characteristics such as shape, width, and position in frequency, power-handling
and sensitivity can be tuned to the application.  Moreover, the
micro-fabrication techniques used in the production of these thin-film devices mean that SFBSs are
intrinsically low-mass, physically compact and easily reproducible, making them
well-suited for array applications  for both ground-based and satellite-borne instruments.
High signal detection efficiencies  can be
achieved up to the superconducting pair-breaking threshold of the
superconductors used in the design; typically 680\,GHz for Nb, but higher
for superconducting compounds such as NbN or NbTiN.

%

Applications for astronomy include  surveys of  moderate
 red-shift galaxies  ($Z=4-6$)  by precision determination of the frequencies of
CO and $[\textrm{CII}]$ rotational lines,
multichroic pixels for cosmic microwave background (CMB) observations
(foreground subtraction by observing in the $24-30$\cite{Koopman_ACTpol_2018} and
$30-48\,\rm{GHz}$ atmospheric windows,
and simultaneous observation in
multiple CMB frequency bands\cite{Pan_SPT3G_2018,Stebor_Simons_TES_spec})
or observation of low-Z CO and O line emissions from nearby galaxies.\cite{Grimes_Camels}
Chip spectrometers coupling to superconducting kinetic inductance detectors (KIDs)
are being developed by a number of groups
and large multiplexing counts of order $1000$'s have
been demonstrated.\cite{Endo_Deshima_2019,Redford_Superspec_2019}
However KID detectors are difficult to design for detection at low frequencies
$\nu \lesssim 100\,\rm{GHz}$, the pair-breaking threshold for
a typical low superconducting  transition temperature material such as Al,\cite{Songyuan_2018a}
and can also be challenging to calibrate as regards their power-to-output-signal responsivity.
Superconducting transition edge sensors  (TESs) are a type of bolometric detector
where the power absorption is
not limited by the pair-breaking threshold of a superconducting film.
In addition their (power-to-current) responsivity  is straightforward to determine both theoretically and experimentally.
%
TES multiplexing schemes are a mature technology using both time- and frequency-domain approaches giving
multiplexing factors of order $100 $'s,\cite{Pourya_fdm_2018, Henderson_tdiv_ACTpol}
and microwave TES readout schemes promise to equal the
multiplexing factors demonstrated for KIDs.\cite{Henderson_microwave_readout, Irwin_microwave_readout}

Vertical profiles of atmospheric temperature and humidity measured
by satellite-borne radiometric sounders provide vital information for long-range weather forecasting.
These sounders work by measuring the upwelling radiance from the
atmosphere in a number of spectral channels, typically either at microwave (MW) or infrared (IR) wavelengths.
Vertical resolution and measurement accuracy improve rapidly with
greater channel number and radiometric sensitivity.\cite{aires2015microwave,Prateek_Hymas}
Significant progress has been made in IR sounder performance;
the Infrared Sounding Interferometer (IASI), for example, provides
over eight thousand channels with sub-kelvin noise equivalent differential temperature (NETD).\cite{iasi}
However, while able to provide high quality data,
IR sounders can do so only under infrequent clear-sky conditions, as clouds absorb and interfere with the signal of interest.
MW sounders, by contrast, are not affected by cloud cover, but their use has been hampered by poorer instrument performance.
Channel number is a significant problem: the Advanced Microwave Sounding Unit-A (AMSU-A)
in current use has, for example, only fifteen
 channels,\cite{airs2000algorithm} while the
 planned Microwave Imaging Instrument (MWI) will offer twenty-three.\cite{alberti2012two}
Sensitivity is also an issue and a recent study by Dongre\,\cite{Prateek_Hymas}
has indicated that maintaining and/or improving sensitivity as channel count increases is vital.
In this paper we report on an SFBS with TES readout
as a technology for realising a MW sounding instrument with several hundred channels and sky-noise limited performance.
This would represent a disruptive advance in the field,
allowing measurements of comparable performance to IR sounders under all sky conditions.

The chip spectrometer reported here is a demonstrator for an atmospheric temperature
and humidity sounder
(HYper-spectra Microwave Atmospheric Sounding: HYMAS),  that is being developed to
operate at frequencies
in the range $\nu= 45 - 190 \,\,\rm{GHz}$.\cite{Prateek_Hymas,Hargrave_Hymas} The
demonstrator  was designed to cover the very important
 O$_2$ absorption band at $\nu= 50 - 60 \,\,\rm{GHz}$ for
 atmospheric temperature sounding.
We believe, however, that our prototype designs and initial characterizations  are already
relevant across the broad band of scientifically important research areas described above.

In Sec.~\ref{sec:tech_overview}
we  give a brief overview of SFBSs and the particular features of the
technology that make them attractive for this application.
We will then describe the design of a set of devices  to
 demonstrate the key technologies required for temperature sounding using the O$_2$ absorption band at
  $50 - 60\,\textrm{GHz}$.
In Sec.~\ref{sect:Fab}
we describe the fabrication of the demonstrator
 chips and their assembly with electronics into a waveguide-coupled detector package.
In Sec.~\ref{sect:results} we
report the first characterization of complete  SFBS's with TES readout detecting at
$40 - 65\,\textrm{GHz}$, considering
the TES response calibration, measurements of overall detection efficiency, and measurements of filter response.
 Finally we   summarise the achievements and
describe our future programme and
the pathway from this demonstrator to a full instrument.

\section{Superconducting on-chip Filter-bank Spectrometers}\label{sec:tech_overview}
\begin{figure}[!ht]
\begin{center}
\begin{tabular}{c}
\includegraphics[width=8cm]{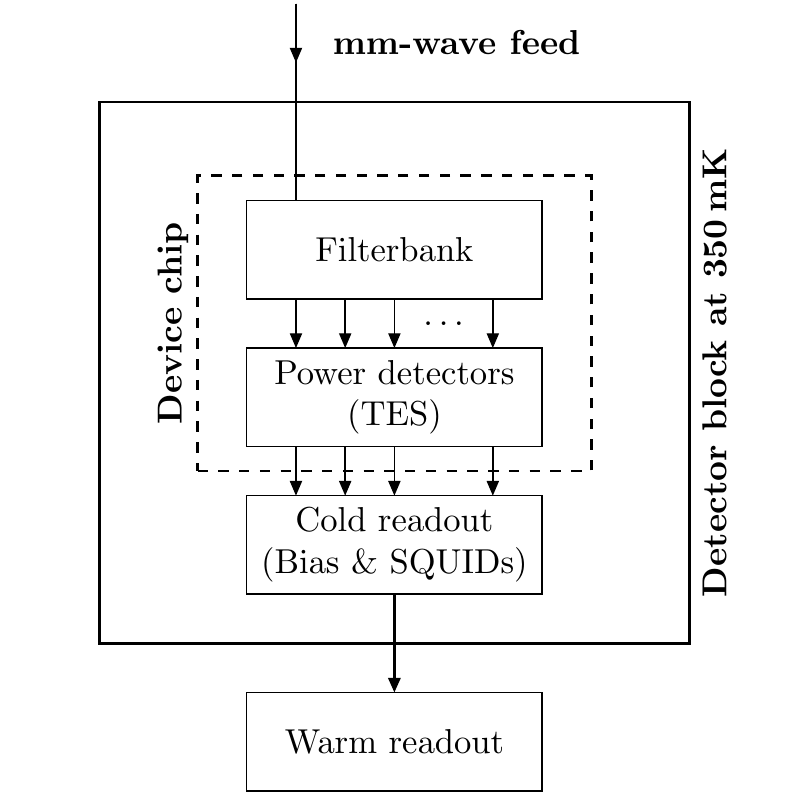}
\end{tabular}
\end{center}
	\caption{\label{fig:system_diagram} Conceptual design of the test devices,
also showing the architecture of a superconducting on-chip filter-bank spectrometer.}
\end{figure}
\begin{table}[b]
\caption{\label{tab:demonstrator_performance} Table of
detector specifications for a baseline temperature sounder using the O$_2$ absorption band.
	The background power loading and detector time constant have been calculated
assuming the same optics and scan pattern as the MWI instrument.\cite{alberti2012two}
	The required NEP gives near  sky-noise limited performance under these conditions.}%
\begin{ruledtabular}
\begin{tabular}{lcl}

	Parameter & Specification & Units \\
\colrule
        & & \\
		Operating frequency range & 50--60 & GHz \\
		Filter resolution & 100-500 & \\
		Noise equivalent power  & 3  & $\textrm{aW}/\sqrt {\textrm{Hz}}    $\\
		Detector time constant & 5 &  ms \\
		Detector absorption efficiency & $50\%$ &  \\
		Background power handling & 60 & fW \\
		Operating temperature & $>300 $ & mK \\
		Number of channels & 35 & \\
	\\
\end{tabular}
\end{ruledtabular}
\end{table}

In general, a filter-bank spectrometer uses a set of band-defining electrical filters to disperse the different spectral components of the input signal over a set of power detectors.
In the case of an SFBS, the filters and detectors are implemented using microfabricated superconducting components,
 and are integrated together on the same device substrate (the `chip').
Integration of the components on the same
chip eliminates housings, mechanical interfaces and
losses between different components, helping to reduce
 the size of the system, while improving ruggedness and optical efficiency.
In addition it is easy to replicate a chip and therefore a whole
spectrometer, making SFBSs an ideal technology for realising spectroscopic imaging arrays.

Most mm- and sub-mm wave SFBSs that have been reported in the literature operate directly at
the signal frequency, i.e. there is no frequency down-conversion step.\cite{Endo_Deshima_2019,Redford_Superspec_2019}
There are two main benefits of this approach, the first of which is miniaturization.
The size of a distributed-element filter is intrinsically inversely proportional to the
frequency of operation: the higher the frequency, the smaller the individual filters and
the more channels that can be fitted on the chip.
The second benefit is in terms of instantaneous observing bandwidth, which is
principle limited only by the feed for frequencies below the pair-breaking thresholds of the superconductors.

Operation at the signal frequency is made possible by: (a) the
availability of superconducting detectors for mm- and
sub-mm wavelengths, and (b) the low intrinsic Ohmic loss of the superconductors.
Critically, (b) allows filter channels with scientifically useful resolution to be realised at high frequencies.
In the case of normal metals, increases in Ohmic loss with frequency and miniaturisation of the
components quickly degrade performance.
As an example, the Ohmic losses in Nb microstrip line at sub-kelvin
temperatures are expected to be negligible for frequencies up
to 680\,GHz\,\cite{yassin1995electromagnetic} (the onset of pair-breaking).
The use of ultra-low noise superconductor detector technology such as TESs and KIDs, (a),
in principle allows for extremely high channel sensitivities.

The SFBS test chips described here were developed with the target application of
satellite atmospheric sounding and against the specification given in Table 1.
Figure 1 shows the system level design of the chips.
Each chip comprises a feeding structure that couples signal from a waveguide onto
the chip in the range 45--65\,GHz, followed by a filter-bank and a set of TES detectors.
The chip is then housed in a test fixture
that incorporates additional cold electronics and waveguide interfaces.
In the sub-sections that follow we will describe the design of each of the components in detail.

Each of the test chips described has twelve detectors in total.
This number was chosen for convenient characterisation without
multiplexed readout, but the architecture readily scales to higher channel count.
In principle the limiting factor is increasing attenuation on the
feed line as more channels are added, but the losses on superconducting line are so low that other issues, such as readout capacity, are expected to be the main limit.

\subsection{Feed and Transition}\label{sec:feed_and_trans}

\begin{figure}[htp]
\begin{center}
\begin{tabular}{c}
\includegraphics[width=8cm]{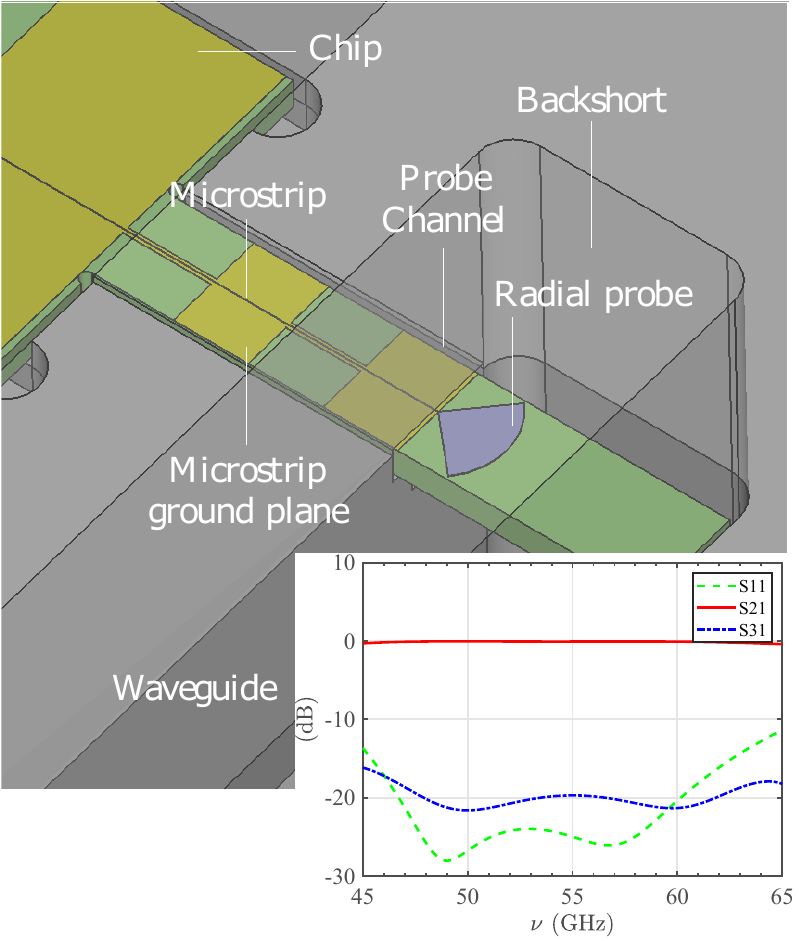}
\end{tabular}
\end{center}
\caption{\label{fig:transition}  Details of the waveguide to microstrip transition.
The different components are described in the text.
In the inset shows the simulated scattering parameters where
 ports 1, 2 and 3 are the waveguide input, microstrip output and coupling the chip-cavity, respectively.
}
\end{figure}

RF signal  input to the test devices is via a standard 1.88\,mm$\times$3.76\,mm WR15 waveguide,
allowing full coverage of the O$_2$ band
(WR15 has a recommended operating range of  $50 - 75\,\textrm{GHz}$, single-mode operation from
$39.875 - 79.750\,\textrm{GHz}$).
The signal is coupled from the waveguide onto a 22.3\,$\Omega$
microstrip line on the chip through a radial probe transition,\cite{kooi2003full}
the design of which is shown in Fig. \ref{fig:transition}.
A split block waveguide is used, the upper part of which has been
rendered as transparent in the figure to allow the internal structure to be seen.
A channel is machined through the waveguide wall at the split
plane to accommodate a silicon beam extending from the chip, shown in green.
This beam supports a fan-shaped radial probe, shown in blue,
 the apex of which connects to the upper conductor of the microstrip.
The microstrip ground plane is shown in yellow.
Not visible is an air channel under the beam, which raises
the cut-off frequency of the waveguide modes of the loaded channel above the band of operation.

It is critical for performance that the ground plane is at the
same potential as the waveguide/probe-channel walls at the probe apex,
however it is not straightforward to make a physical connection.
The changes in ground plane width shown in Fig. \ref{fig:transition} implement
a stepped impedance filter in the ground-plane/wall system to ensure a wideband
short at the probe plane, assuming the ground plane is wire-bonded to the walls in the chip cavity.
This filter also prevents power flow along the air channel due to the TEM mode.

The performance of the  design was simulated using OpenEMS,
a finite difference time domain (FDTD) solver.\cite{openEMS}
Insertion loss ($S_{21}$), reflection loss ($S_{11}$) and indirect
 leakage to the chip cavity ($S_{31}$) as a function of frequency are shown in the inset of Fig. \ref{fig:transition}.
As can be seen, the design achieves better than -3\,dB
insertion loss over $45 - 65\,\textrm{GHz}$, corresponding to a fractional bandwidth of nearly 35\,\%.

\subsection{Filter-bank}\label{sec:filter-bank}


\begin{figure}
\begin{subfigure}{\linewidth}
\centering
\includegraphics[width=6.5 cm]{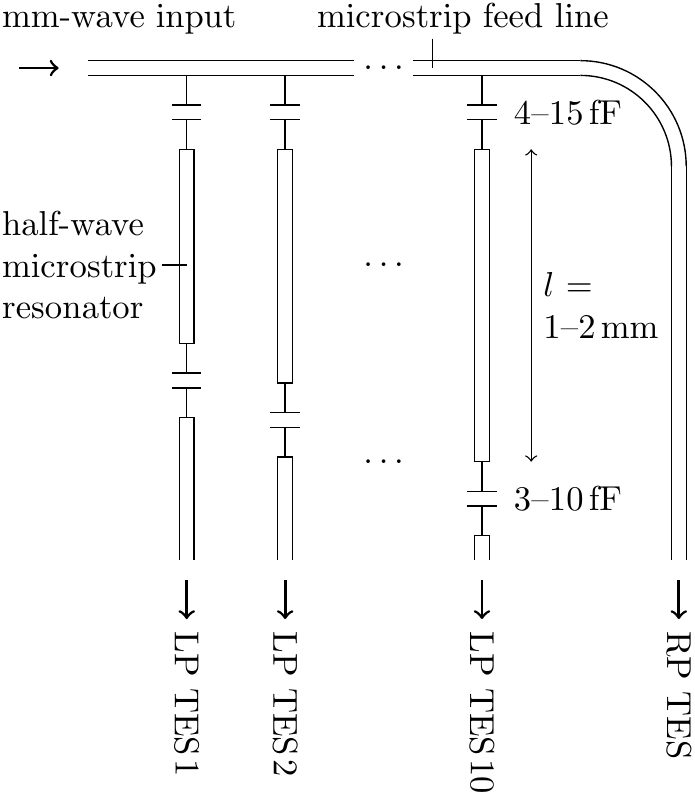}
\caption{\label{fig:filter-bank_schematic} Schematic of the filter-bank.}
\end{subfigure} \\
\vspace{0.5 cm}
\begin{subfigure}{\linewidth}
\centering
\includegraphics[width=7.5cm]{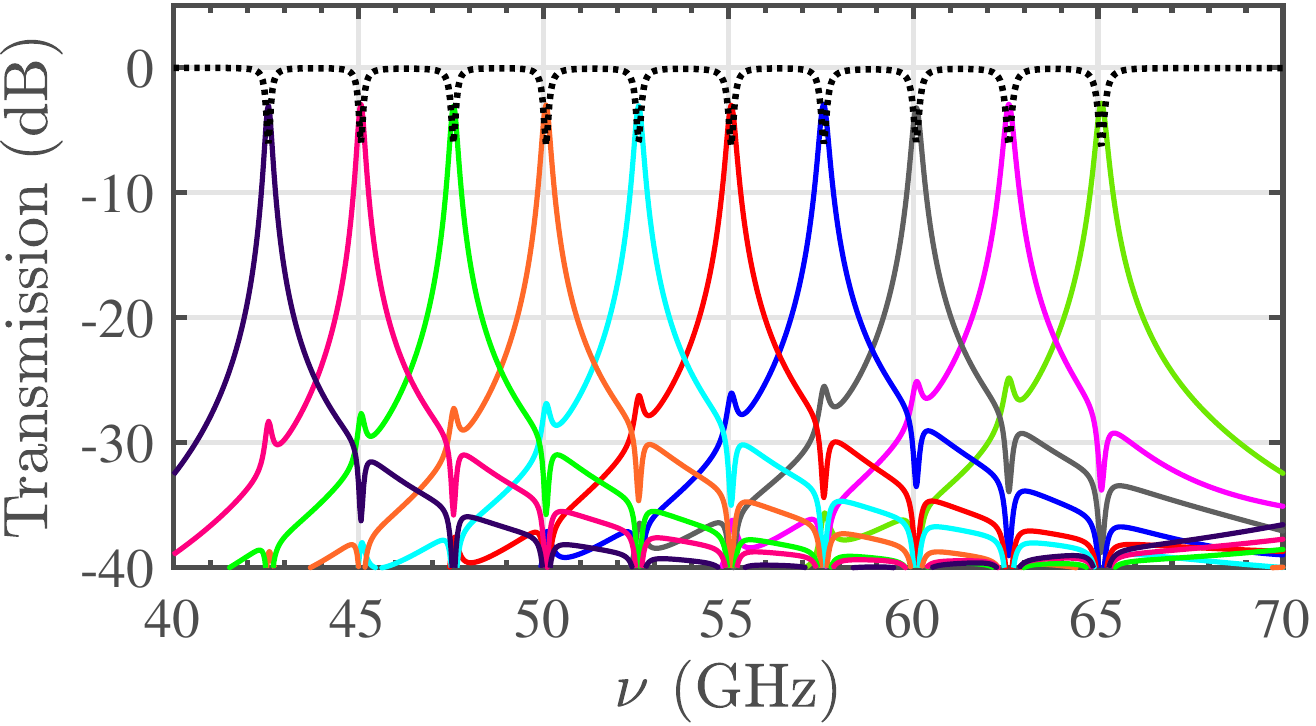}
\caption{\label{fig:filter-bank_performance} Modelled transmission gain versus frequency.
}
\end{subfigure}
\caption{\label{fig:filter-bank} Details of the filter bank designs.
Fig \ref{fig:filter-bank_performance} shows the power transmitted to
individual channels for $\mathcal{R}=200$. The different channels are indicated by the different line colours.
Transmission to the residual power detector terminating the feed is indicated by the black dotted line.
}
\end{figure}

The filter-bank architecture employed on the test chips is shown schematically in Fig. \ref{fig:filter-bank_schematic}.
It comprises ten half-wave resonator filters driven in parallel from a single microstrip feed line.
Each of the filters comprises a length of microstrip line,
the ends of which are connected by overlap capacitors to the feed and an
output microstrip line. The output line  is terminated by a matched resistor that is closely thermally-coupled
to a TES bolometer.
Both the termination resistor and TES are thermally isolated from the cold stage.
The feed line itself is terminated on an eleventh
matched resistor coupled to a TES.
This TES measures the power remaining after passing the filter bank
and is subsequently described  as the residual power detector.
This TES was specifically designed to handle the expected higher incident power.
A twelfth detector on chip has  no microwave connection and was used as a `dark' (i.e. nominally power-free)
reference.
A common superconducting microstrip design is used for all parts of the filter-bank, consisting of a Nb ground plane,
a 400\,nm thick SiO$_2$ dielectric layer and a 3\,$\mu$m wide, 500\,nm thick, Nb trace.
The modelled characteristic impedance is 22.3\,$\Omega$.\cite{yassin1995electromagnetic}

The spectral characteristics of each filter are determined by the resonant behaviour of the line section
and the strength of the coupling to the feed and detector.
For weak coupling, the line section behaves as an open-ended
resonator and transmission peaks sharply at resonance.

Each filter channel
is designed to have its own fundamental frequency $\nu_0$.
For a filter of length $l$
 and phase velocity $c$,  $\nu_0$ is given by
\begin{equation}\label{eqn:resonant_frequency}
	\nu_0 = \frac{c}{2 l}.
\end{equation}
Tuning over the range $40-70\,\textrm{GHz}$ is achieved by values of $l$ in the range $2.225 - 1.00\,\textrm{mm}$.
The channel resolution $\mathcal{R}=\nu_0/\Delta \nu$ where $\Delta \nu$ is the 3-dB width.

The bandwidth of a filter and its peak-value of transmission
are governed by losses in the resonator, as measured by the quality factor.
Assuming power loss $P$ from the resonator when the stored energy is $U$,
the total quality factor $Q_\text{t}$  of the resonator
(and hence the  resolution of the filter $\mathcal{R}\equiv Q_\text{t}$) is defined
by $Q_\text{t} = 2 \pi \nu_0 U / P$.
$P$ can be further decomposed into the sum of the power loss $P_\text{c,in}$
and $P_\text{c,out}$ to the input and output circuit respectively (\emph{coupling losses})
and the power dissipated $P_\text{int}$ in the resonator itself through Ohmic and dielectric loss (\emph{internal losses}).
We can then define additional quality factors $Q_\text{int}$, $Q_\text{c,in}$ and $Q_\text{c,out}$
by $Q_{n} = 2 \pi \nu_0 U / P_n$
These correspond to the Q-factor of the resonator in
the limit where the individual losses are dominant,
and $Q_\text{t}^{-1} = Q_\text{int}^{-1} + Q_\text{c,in}^{-1} + Q_\text{c,out}^{-1}$.
With these definitions, the power gain of the channel as a function of frequency $\nu$ can be shown to be
\begin{equation}\label{eqn:filter_power_gain}
	G(\nu,\nu_0) = \frac{2 Q_\text{t}^2}{Q_\text{c,in} Q_\text{c,out}}
	\frac{1}{1 + 4 Q_\text{t}^2 (\nu - \nu_0)^2 / \nu_0^2}.
\end{equation}
%
$Q_\text{c,in}$ and $Q_\text{c,out}$ are controlled by the input and output coupling capacitors.
For maximum transmission of power to the filter channel detector (-3\,dB) we must
engineer $Q_\text{c,in} = Q_\text{c,out} = Q_\text{c} \ll Q_\text{int}$,
in which case $Q_\text{t} = Q_\text{c} / 2 \ll Q_\text{int}$.
Therefore the minimum achievable bandwidth is limited by $Q_\text{int}$.
Under the same conditions the power transmitted  to the residual power detector is
\begin{equation}\label{eqn:filter_power_transmitted}
	T(\nu,\nu_0) = 1+ \frac{Q_\text{t}}{Q_\text{c,in}} \frac{\left[ Q_\text{t}/Q_\text{c,in} -2 \right]}
{1 + 4 Q_\text{t}^2 (\nu - \nu_0)^2 / \nu_0^2},
\end{equation}
and the power is reduced by 6-dB on resonance as
 seen from the dotted black line in Fig.~\ref{fig:filter-bank_performance}.
In practice the presence of the coupling capacitors causes the measured centre frequency $\nu_\text{0,meas}$ of the filter to differ slightly from the resonant frequency of the open-ended line, as given by (1).
The detuning is dependent on coupling strength and analysis of the perturbation of the circuit to first-order gives
\begin{equation}\label{eqn:detuning}
	\nu_\text{0,meas} \approx
	\left( 1 - \frac{1  + \sqrt{2}}{2 \sqrt{\pi \mathcal{R}}} \right) \nu_0 .
\end{equation}

Ohmic loss is small in superconducting microstrip lines at low temperatures.
Instead, experience with superconducting microresonators $<$10\,GHz suggests
dielectric loss due to two-level-systems will govern $Q_\text{int}$. \cite{zmuidzinas2012superconducting}
A main aim of the prototype  devices was to investigate the loss mechanisms active
at millimetre wavelengths, although extrapolation from the low-frequency
data suggests spectral resolutions
 of a few hundred should be easily achievable for the microstrip design used here, and higher
may be achievable for alternatives such as coplanar waveguide.

We report on two of the four different filter bank designs that were fabricated on the demonstrator devices.
Three of these have identically spaced channels, designed to be 2.5\,GHz apart over the range $42.5 - 65.00\,\textrm{GHz}$.
The designs differ in spectral resolution of the channels, with target values of $\mathcal{R}=$250, 500 and 1000.
These explore the achievable control of channel placement and resolution
and provide experimental characterization of the millimetre-wave properties of the microstrip over a wide frequency range.
However, the channels in these designs are widely spaced.
The fourth design investigated an alternate possible mode of operation where the passbands of the
filters overlap significantly, with nine $\mathcal{R}=500$ channels in the band $53 - 54\,\textrm{GHz}$.
In this case the spacing of the filters along the feed line becomes significant, as the
filters interact with each other electrically.
If the filters are arranged in order of decreasing frequency and placed approximately a quarter of a
wavelength apart, part of the in-band signal scattered onto the feed by the filter is reflected back
in phase from the next filter along, enhancing transmission to the detector.
Consequently the transmission of the filters can be increased by dense packing,
but at the expense of the shape of the response; we are investigating the implications for scientific performance.
The modelled passbands for the widely-spaced filter banks are shown in Fig. \ref{fig:filter-bank_performance}.

\subsection{TES design}
\label{sect:TES_Design}
Electrothermal feedback (ETF) in a thermally-isolated TES of resistance $R$,
voltage-biased within its superconducting-normal
resistive transition
means that the TES
self-regulates its  temperature $T$  very close to $T_c$, the TES transition temperature.\cite{Irwin2005}
When the superconducting-normal resistive transition
occurs in a very narrow temperature range  $\alpha\gg 1$, where  $\alpha=T (dR/dT)/R$
characterizes the transition sharpness.
Provided
$T,\,T_c\gtrapprox 1.5 T_\textrm{b}$ the bath temperature, the small-signal power-to-current responsivity
$s_I$
is then given by $s_I=-1/(I_0(R_0-R_\textrm{L})$. Here $T_0$, $I_0$ and $R_0$ are the TES temperature, current and resistance
respectively at the operating point  and $R_\textrm{L}$ is the load resistance.
All are simple to evaluate from  measurements of the TES and its known bias circuit parameters
meaning that  $s_I$ is straightforward to evaluate.

The TESs we report on were designed to operate from a bath temperature of $T_\textrm{b}\simeq300\,\,\rm{mK}$
in order  to be usable from a simple cryogenic cooling platform, to have a saturation power of order
$P_\textrm{sat}\simeq 2\,\,\rm{pW}$ to satisfy expected power loading, and a
 phonon-limited noise equivalent power (NEP) of order $2\,\rm{aW/\sqrt{Hz}}$
to minimise detector NEP with respect to atmospheric noise (see Table~\ref{tab:demonstrator_performance}).
TES design modelling indicates that the required performance
 should be achievable with a superconducting-normal transition temperature
$T_c\sim 550\to 650\,\rm {mK}$. The detailed calculation depends on the value of the
exponent $n$  that occurs in the calculation of the
power flow from the TES to the heat bath: $P_\textrm{b}=K_\textrm{b}(T_c^n-T_\textrm{b}^n)$,
where $K_\textrm{b}$ is a material dependent parameter that includes the
geometry of the thermal link between the TES and bath. For ideal voltage bias
($ R_\textrm{L} =0 $)
the saturation power
is  $P_{\textrm{sat}}=(1-R_0/R_{\textrm{N}})P_\textrm{b}$
and $R_{\textrm{N}}$ is the TES normal-state resistance.\cite{Irwin2005}
Under typical operating conditions $R_0\sim 0.25 R_{\textrm{N}}$.
 The thermal conductance to the bath $G_\textrm{b}=dP_\textrm{b}/dT_c$, determines the phonon-limited
 NEP where $\textrm{NEP}=\sqrt{4k_\textrm{b}\gamma G_\textrm{b} T_c^2}$,
  $k_\textrm{b}$ is Boltzmann's constant, and $\gamma\leq 1$
 takes account of the temperature gradient across $G_\textrm{b}$. Our previous   work suggests that
 at the operating temperature and length scales  $n\sim 1.5\to 2.5$
 should apply for 200-nm thick \SiN\, at these temperatures with
 lengths $50-1000\,\rm{\mu m}$. For the filter channels, thermal
 isolation of the TES was formed
 from four 200nm-thick \SiN\, legs each of length $500\,\rm{\mu m}$,
 three of width $1.5\,\rm{\mu m}$ and one of
 $4\,\rm{\mu m}$ to carry the microstrip.
 The residual power detector had support legs of length $50\,\rm{\mu m}$ and the same widths as for the filter channels.

\section{Fabrication and Assembly}
\label{sect:Fab}

\begin{figure}
\begin{subfigure}{\linewidth}
\centering
\includegraphics[trim = {0.35cm 0cm 0cm 0cm}, clip, width=8cm]{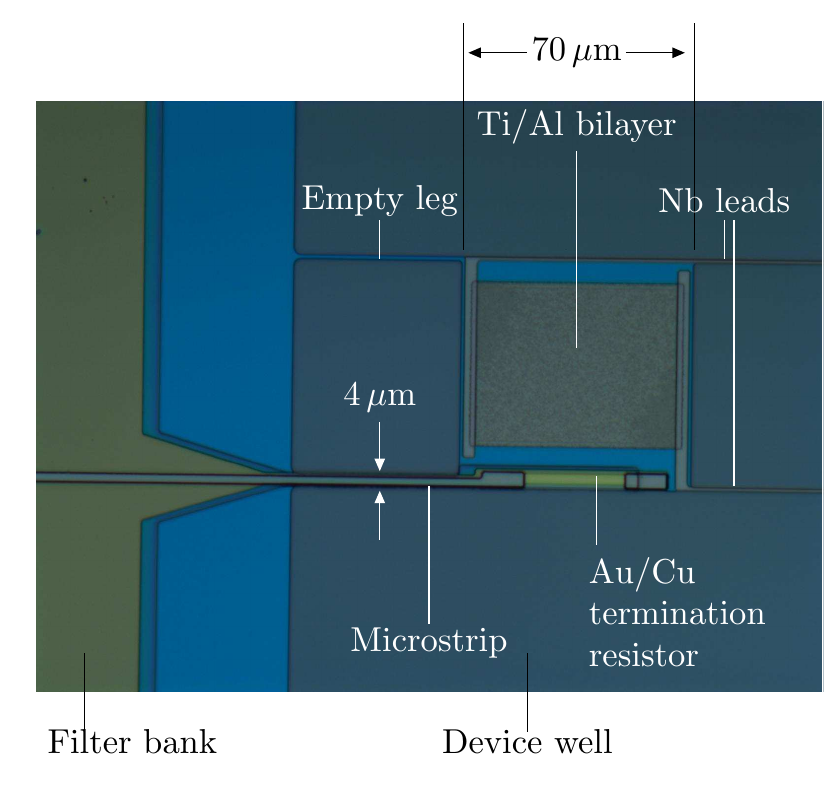}
\caption{\label{fig:detector_photo} High resolution optical image of a  residual power detector.}
\end{subfigure} \\
\begin{subfigure}{\linewidth}
\centering
\includegraphics[trim = {0.35cm 0cm 0cm 0cm}, clip, width=8cm]{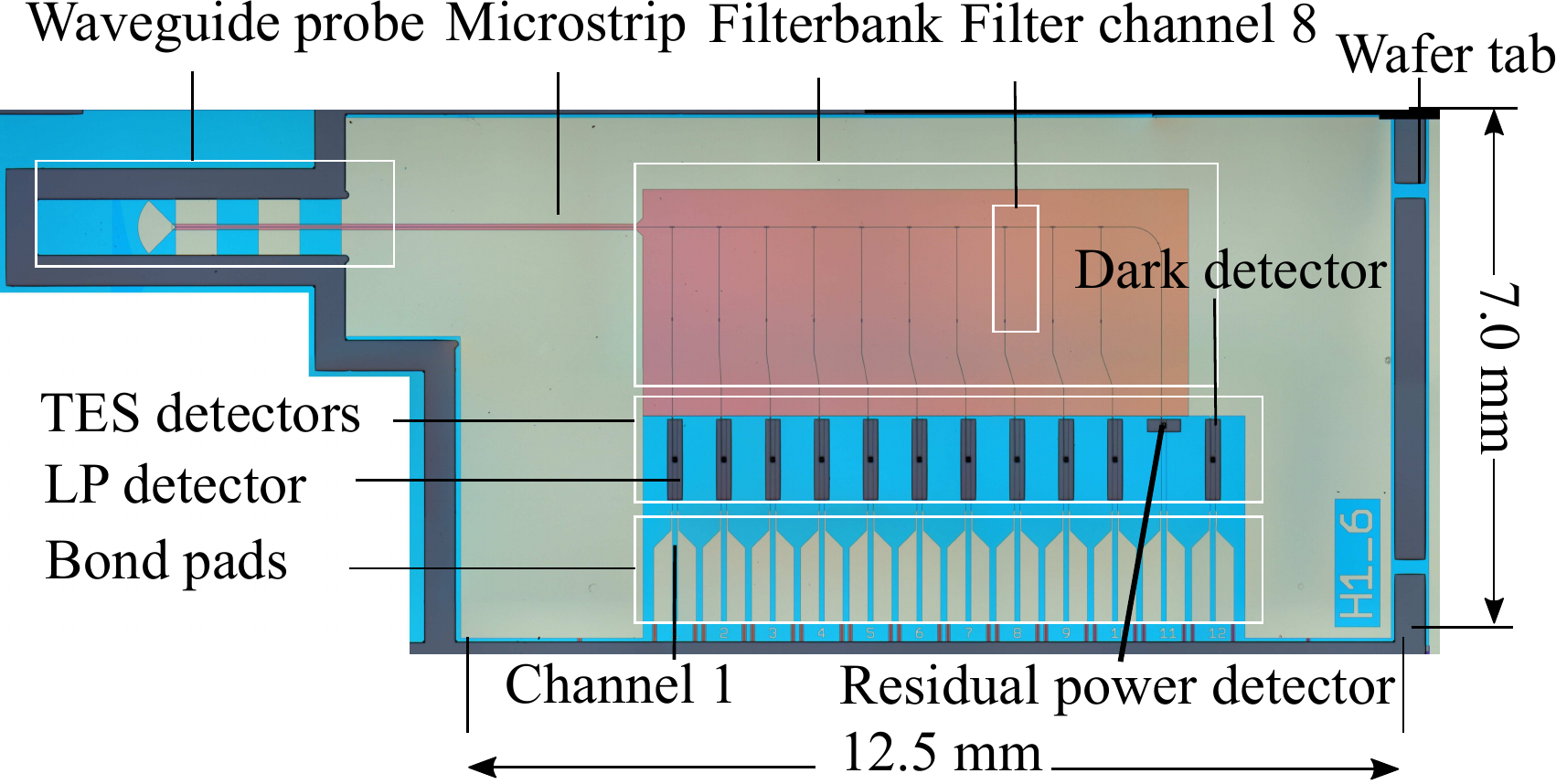}
\caption{\label{fig:chip_photo} Optical image of a complete demonstrator chip. The contact pads for Channel 1
 are
indicated in the lower left of the image, with channel numbers increasing to the right. }
\end{subfigure}
\caption{\label{fig:detectors} Photographs of the residual power detector and a complete demonstrator chip.
Both images were taken prior to DRIE of the device wells, that removes the dark-grey areas to release the membranes
and the chip itself.}
\end{figure}

The detector chips were fabricated on $225\,\rm{\mu m}$-thick Si wafers coated with
a dielectric bilayer comprising 200~nm thick low-stress \SiN\,\,
and an additional 50~nm
\SiO\,
etch stop. After Deep Reactive Ion Etching (DRIE) the \SiN, \SiO\,  films formed the thermal isolation structure.
The detectors were made by
sputtering metal and dielectric thin films
 in ultra-high vacuum.
The superconducting microstrip was formed from a 150~nm-thick Nb ground plane
with 400~nm amorphous silicon dioxide (\SiO)
dielectric and a 400~nm-thick $3\,\rm{\mu m}$-wide Nb line.
Coupling capacitors were made from \SiO\,
and Nb.
Thin film AuCu resistors terminate the microstrip  on the TES island.
The TESs
were fabricated from TiAl bilayers with the Ti and Al thicknesses
of
thicknesses 150~nm, 40~nm respectively, calculated to target a superconducting transition temperature of
 $600-650\,\rm{mK}$.\cite{Songyuan_Tc_2018}
  Electrical connections to the TES were sputtered Nb.
  DRIE
 removed the Si under the TES such that the \SiN\,
 island and legs provide the necessary thermal isolation.
 The DRIE also released the individual chips from the host wafer and defined the chip shape
 seen in \ref{fig:detectors}.

Figure~\ref{fig:detector_photo} shows an image of the residual-power detector prior to DRIE of the Si.
 The $4\,\rm{\mu m}$-wide
leg supporting the microstrip is indicated in the lower left-hand region of the image.
The $3\times35\,\rm{\mu m}$ AuCu termination
can be seen.
Figure~\ref{fig:detectors}~(b) shows a composite high resolution photograph of a completed detector chip after DRIE.
The component structures, waveguide probe,
filter-bank, TES detectors and superconducting contact pads are indicated.
The individual filter channels are also
visible.
Filter lengths are  determined by the small coupling capacitors that are (just) visible as darker dots between the
through-line and the individual TES wells, the lengths increase with
channel number (left-to-right in the image) and correspondingly  $\nu_{0}$ reduces.

\begin{figure}[htp]
\begin{center}
\begin{tabular}{c}
\includegraphics[width=8.0cm]{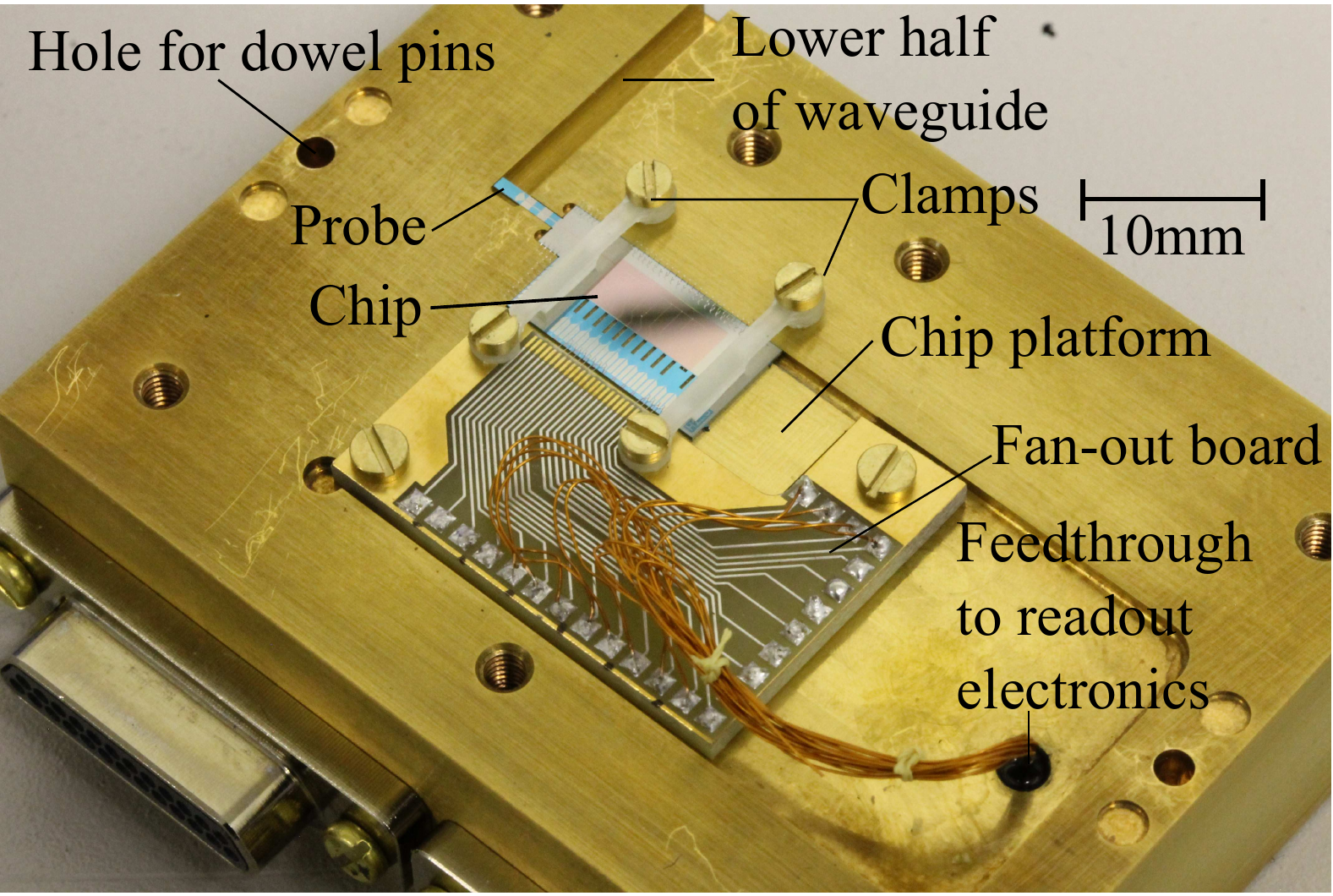}
\end{tabular}
\end{center}
\caption
{ \label{fig:Detector_cavity}
Photograph of the assembled lower half of the detector enclosure. The upper part forms a light tight enclosure and completes
the upper section of the split waveguide. }
\end{figure}
Figure~\ref{fig:Detector_cavity} shows an image of a completed chip mounted in a detector enclosure. The probe can
be seen extending into the lower half of the split waveguide. Al wirebonds connect the  chip ground plane to the
enclosure and on-chip electrical wiring to the
superconducting fan-out wiring.  Superconducting NbTi wires connect
through a light-tight feed-through into the electronics enclosure
on the back of the detector cavity. The low-noise
two-stage SQUIDs used here were provided by Physikalisch-Technische Bundesanstalt (PTB). The upper section of the detector
enclosure completed the upper section of the waveguide and ensured that the package was light-tight.
Completed detector packages were cooled using two separate cryogen-cooled  ADR's
 both giving a base temperature $T_{\textrm{b}}\lesssim100\,\,\rm{mK}$.

\begin{figure}[htp]
\begin{center}
\begin{tabular}{c}
\includegraphics[width=8cm]{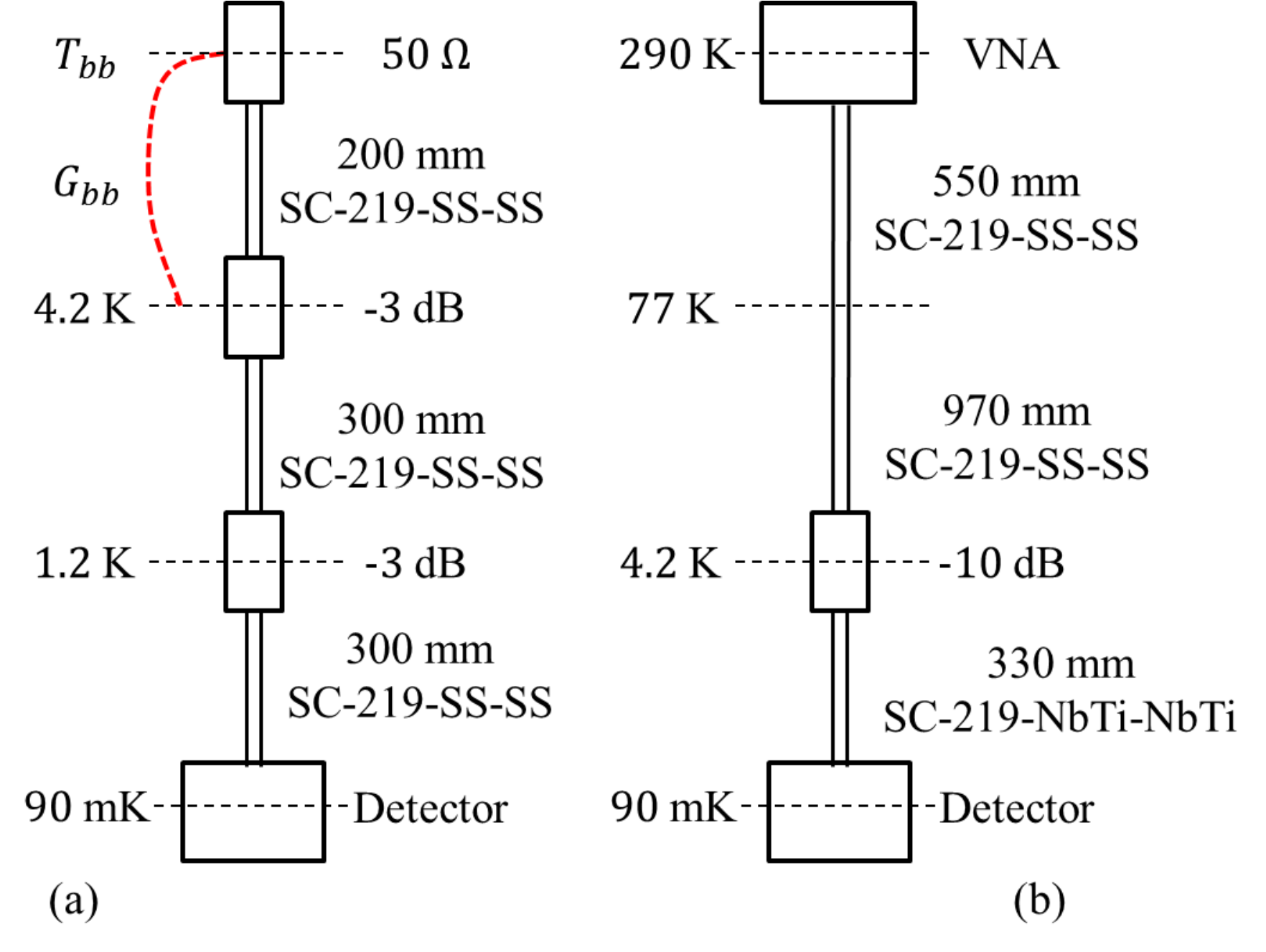}
\end{tabular}
\end{center}
\caption
{ \label{fig:BB_schematic}
Schematics of the measurement scheme. (a) For the blackbody power calibration. (b) for the measurement of filter spectral response}
\end{figure}
Figure~\ref{fig:BB_schematic}~(a) shows a schematic of the blackbody power calibration scheme. A matched $50\,\,\rm{\Omega}$
resistive load (Fairview Microwave SKU:ST6510)\cite{Fairview}
terminated a length of Coax Co. Ltd SC-219/50-SS-SS coaxial line.\cite{Coax_jp}
The load was
mounted on a thermally isolated
Cu plate within the cryostat $4\,\textrm{K}$ vacuum space.
This termination-resistor stage was equipped with a calibrated LakeShore RuOx  thermometer\cite{Lakeshore}
 and resistive heater
and  was connected to the refrigerator $4\,\,\rm{K}$ stage by a length of copper wire ($l=20\,\,\rm{cm}$, $r=250\,\,\rm{\mu m}$)
 that determined the dominant stage time constant which was estimated at $\sim 20\,\,\rm{ms}$.
Further SC-219
connected the termination resistor to a 3-dB attenuator (Fairview Microwave SA6510-03)
 mounted on the
 $4\,\,\rm{K}$ stage itself, 
 to a further attenuator mounted on the
 $1\,\,\rm{K}$ stage, in order to minimise heat-loading to the cold-stage
 by heat-sinking the inner coax conductor, and then to the WR-15
 coax-waveguide transition (Fairview Microwave SKU:15AC206), waveguide probe
  and detector chip. The total coax length was
 $800\,\,\rm{mm}$. Results of these measurements are reported in Sec.~\ref{sect:Power_calibration}.

The spectral response of the filters was measured using a continuous wave (CW) source.
For these
tests a different cryostat was used, but it again comprised a two-stage (1K/50mK) ADR launched
from a 4\,K plate cooled by liquid cryogens.  A different set cold electronics
with nine single-stage SQUIDs, provided by PTB was used
that allowed simultaneous readout of nine devices.

Figure~\ref{fig:BB_schematic}~(b) shows a schematic of the scheme.
A Rohde \& Schwarz ZVA-67 VNA capable of
measurements up to $67\,\textrm{GHz}$ was used as a frequency-tuneable, power-levelled, CW source and the
signal coupled down to the detector block on coaxial cable.  A KMCO KPC185FF HA hermetic
feedthrough was used to enter the cryostat, then 1520~mm of Coax Co. Ltd SC-219/50-SS-SS coaxial
cable was used for the connection between room temperature and the $4\,\textrm{K}$ plate, anchoring on the liquid nitrogen tank,
liquid helium tank and finally the $4\,\textrm{K}$ plate itself.  The final connection from $4\,\textrm{K}$ to the
waveguide adapter on the test block was made using 330\,mm of Coax Co. Ltd SC-219/50-NbTi-NbTi
superconducting cable for thermal isolation.  A 10\,dB attenuator (Fairview Microwave SA6510-10) was inserted in line
at the $4\,\textrm{K}$ plate to provide  thermal anchoring between the inner and outer conductors
of the coaxial cable and isolation of the detectors chip from short-wavelength power loading.
All connections were made with 1.85~mm connectors to allow mode free
operation up to $65\,\textrm{GHz}$.  Measurements indicated the expected total attenuation of order 68\,dB between
room temperature and the WR-15 waveguide flange. In operation the frequency of the CW source was stepped
through a series of values and the output of the devices logged in the dwell time between transitions,
exploiting the normal mode of operation of the VNA.
Results of these measurements are reported in
Sec.~\ref{sect:Filter_response}.

Two chips were characterized in this series of tests. Both chips were fabricated on the same wafer, but have different
designed filter resolving powers. Chip 1 has filters designed with $\mathcal{R}=500$ was used for the blackbody
measurements with measurements on filter channels
4-6, 11 (the residual power detector), and 12 (the dark detector).
Chip 2 with $\mathcal{R}=200$ was used for the spectral measurements
and we report measurements on channels 3-8, 10 and 12.
 Both chips were designed to cover
the  $42.5-65\,\rm{GHz}$ range.

\section{Results}
\label{sect:results}
\subsection{TES characteristics}
\label{sect:TES_results}
\begin{table}[htp]
\caption{\label{tab:TES_dc_results}%
Summary of DC TES characteristics for Chip 1.
}
\begin{ruledtabular}
\begin{tabular}{cccccl}
\textrm{Channel}&
$R_{\textrm{N}}\,\rm{(\Omega)}$&
$T_c\,\rm{(mK)}$&
$n$&
$G_b\,\rm{(pW/K)}$&
Notes\\
\colrule
&&&&&\\
4 & 2.6 & 457 & $2.0\pm0.1$ & $4.7$ & \textrm{Filter}\\
5 & 2.5  & 452 & $2.0\pm0.1$ &  $\rm{\,\,\,}4.75$ & \textrm{Filter}\\
6 & 2.4 & 454 & $2.0\pm0.1$ & $4.8$ & \textrm{Filter}\\
12 & 2.4  & 455 & $2.0\pm0.1$ & $4.7$ & \textrm{Dark}\\
 &   &  &  &  & \\
11 & 2.4  & 459 & $2.0\pm0.1$ & $65\rm{\,\,\,\,\,\,}$ & \textrm{Residual}\\
 &  &  &  &  & \textrm{Power}\\
\end{tabular}
\end{ruledtabular}
\end{table}
Table~\ref{tab:TES_dc_results} shows measured and calculated values for the DC characteristics of 5 channels from Chip 1.
$T_c$ was close-to but  somewhat lower than designed, and $R_{\textrm{N}}$ was higher.
The critical current density was also reduced from our previously measured values even for
pure Ti films. This may indicate that these TESs demonstrate
the thin-film inter-diffusion effect between Ti-Nb couples recently reported by
Yefremenko~{\textit {et al.}}.\cite{Yefremenko2018_2}
The superconducting-normal resistive transition
with $T_{\textrm{b}}$, as inferred from measurements of the TES Joule power dissipation as a function of
 bias voltage, appeared
somewhat broad. Hence,  $G_b$ was determined from the power dissipated at constant TES resistance $R_0=0.25R_{\textrm{N}}$, close
to the bias voltage used for power measurements. The exponent in the power flow was $n=2.0\pm0.1$ for all TESs.

\subsection{TES ETF parameters and power-to-current sensitivity estimate}
\label{sect:ETF_estimates}
%
\begin{figure}[htp]
\begin{center}
\begin{tabular}{c}
\includegraphics[width=8.0cm]{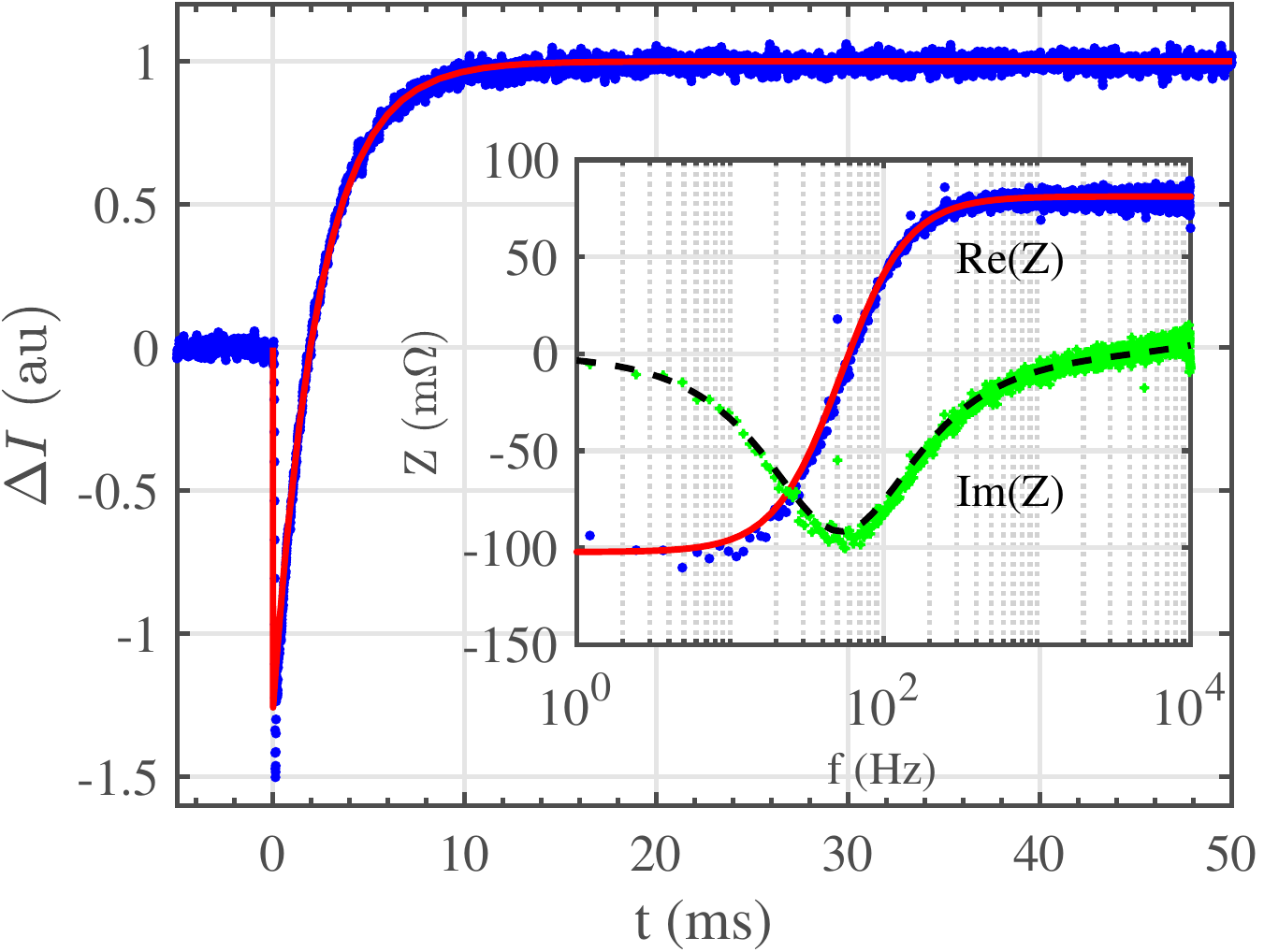}
\end{tabular}
\end{center}
\caption
{ \label{fig:Response_p11_400}
The inset shows the measured real
and imaginary parts (blue and green dots respectively) of the impedance of the residual power detector (channel 11).
The solid red and dashed
black lines are the modelled values (real and imaginary respectively) using the parameters given in Table~\ref{tab:ETF_parameters}.
The main plot shows the
measured (blue dots) and calculated (red line) response of the TES current to a
small step change in the bias voltage for the same channel with no additional parameters.
 }
\end{figure}

For a TES modelled as single heat capacity $C$ with
conductance to the bath $G_\textrm{b}$, the measured impedance $Z(f)$
is given by\cite{Irwin2005}
\begin{equation}
Z(f)=R_\textrm{L}+j2\pi f L_\textrm{in}+Z_\textrm{TES}(f),
\end{equation}
where $R_\textrm{L}$ is the load resistance, $L_\textrm{in}$ is the input inductance of the SQUID plus any stray inductance,
$Z_\textrm{TES}(f)$ is the TES impedance and $f$ is the measurement frequency. The TES impedance is given by
\begin{equation}
Z_\textrm{TES}(f)=R_0(1+\beta)+ \frac{R_0 \mathscr{L}_I}{1-\mathscr{L}_I}\frac{2+\beta}{1+  2 \pi f \tau_I},
\end{equation}
where $\mathscr{L}_I=P_0\alpha/G_\textrm{b}T_0$. $P_0=I_0R_0^2$ is the
Joule power at the bias point, $T_0\simeq T_c$ is the operating temperature and
$\alpha=T\, (dR/dT)/R$ characterizes the sharpness of the resistive transition
at the bias point. $\beta=I (dR/dI)/R$ measures the sensitivity
of the transition to changes in current $I$.  At measurement frequencies much higher than the reciprocal of the effective TES time
 (here taken to be $f\gtrsim 5\,\rm{kHz}$),
 $\textrm{Re}\, ( Z(f))=R_0(1+\beta)$, and so $\beta$ can be determined
with minimal measured parameters.
At low frequencies, here taken to be
 $f\lesssim 5\,\rm{Hz}$, $\textrm{Re}\, ( Z(f)) =R_\textrm{L}+R_0(1+\beta)-(R_0\mathscr{L}_I(2+\beta))/(\mathscr{L}_I-1)$,
 so that $\mathscr{L}_I$ and hence
 $\alpha$ can be found.
The low frequency  power-to-current responsivity is given by
\begin{equation}
s_I(0)= -\frac{1}{I_0R_0}\left[\frac{R_\textrm{L}+R_0(1+\beta)}{R_0\mathscr{L}_I}+1-\frac{R_\textrm{L}}{R_0} \right]^{-1}.
\label{equn:s_I}
\end{equation}
For a TES with good voltage bias $R_\textrm{L}\ll R_0$, and a sharp transition $\alpha, \, \mathscr{L}_I\gg1$, then
 provided $\beta\ll \mathscr{L}_I-1$, $s_I(0)=-1/(I_0(R_0-R_\textrm{L}))$ and the power-to-current responsivity is  straight-forward
 to calculate.
For the TiAl TES reported here the measured
power-voltage response suggested a fairly broad
superconducting-normal transition. Here we use the full expression given by Eq.~\ref{equn:s_I}
with $s_I(0)=-k_s/(I_0(R_0-R_\textrm{L}))$ and $k_s$  the calculated value of the reciprocal within the brackets.

The inset of Fig.~\ref{fig:Response_p11_400} shows the measured real  and  imaginary parts (blue and green dots respectively)
 of the
impedance of the residual power detector, channel 11. The solid red and dashed black lines show the modelled impedance.
Table~\ref{tab:ETF_parameters}
shows the derived ETF  parameters and heat capacity.
The main figure shows the measured normalised TES current response (blue dots) to a small change
in TES bias voltage (i.e. a small modulation of instantaneous Joule power), and (red line) the calculated
response using the  parameters determined in modelling the impedance.
No additional parameter were used.

\begin{figure}[htp]
\begin{center}
\begin{tabular}{c}
\includegraphics[width=8.0cm]{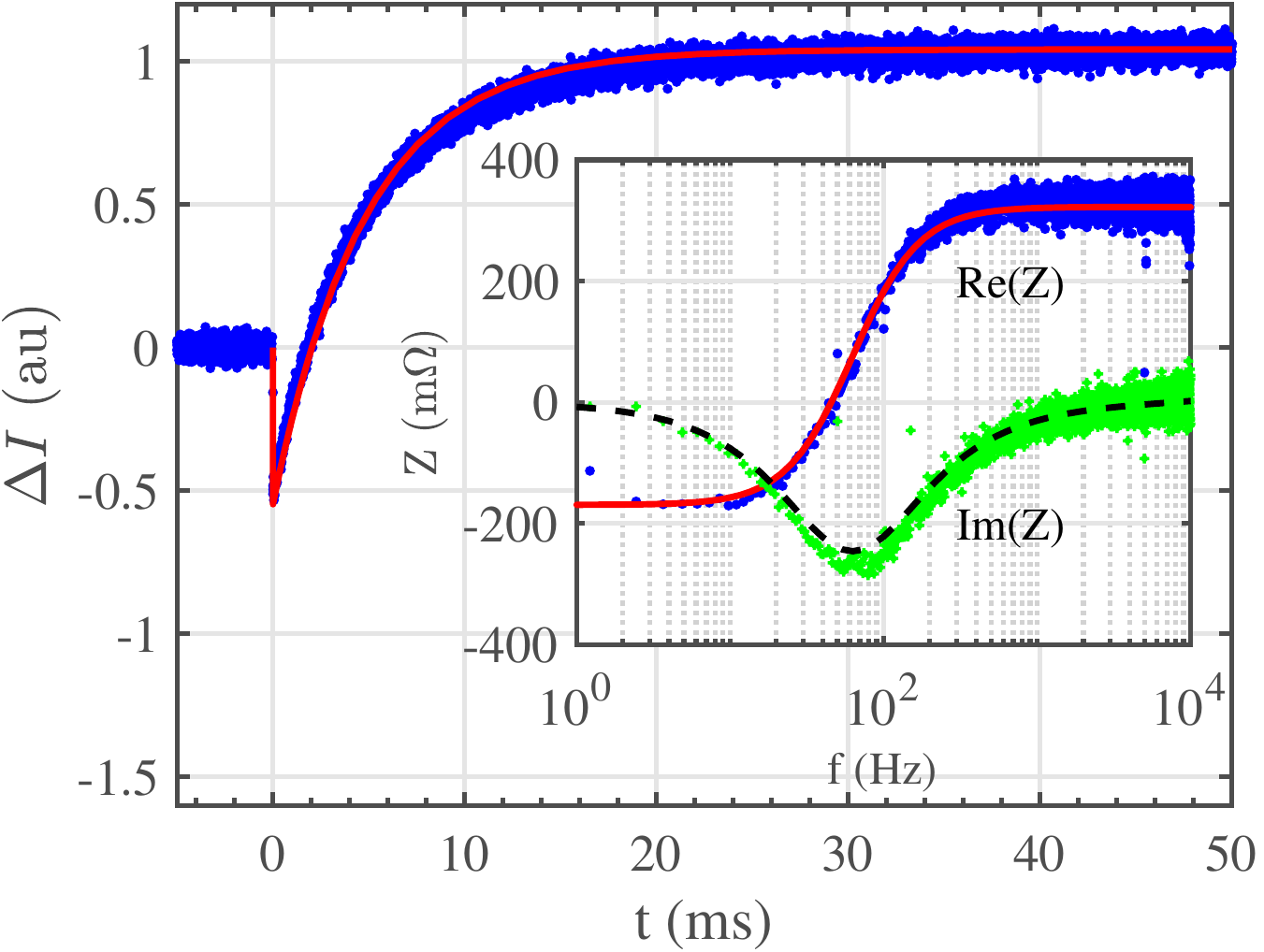}
\end{tabular}
\end{center}
\caption
{ \label{fig:Response_p12_300}
The inset shows the measured real
and imaginary parts (blue and green dots respectively)  of the
impedance of the dark detector (channel 12). The solid red and dashed
black lines are the modelled values (real and imaginary respectively) using the parameters given in Table~\ref{tab:ETF_parameters}.
The main plot shows the
measured (blue dots) and calculated (red line) response of the TES current to a
small step change in the bias voltage for the same channel without additional parameters.
 }
\end{figure}

The inset of Fig.~\ref{fig:Response_p12_300} shows the measured real  and  imaginary parts (blue and green dots respectively)
 of the
impedance of the dark detector, channel 12. The solid red and dashed black lines show the modelled impedance.
Table~\ref{tab:ETF_parameters}
shows the derived ETF  parameters and heat capacity.
The main figure shows the measured normalised TES current response (blue dots) to a small change
in TES bias voltage (i.e. a small modulation of instantaneous Joule power), and (red line) the calculated
response using the  parameters determined in modelling the impedance.
As in the modelling shown in Fig.~\ref{fig:Response_p11_400},
no additional parameter were used.
Comparable correspondence  between
measured impedance and current step response using impedance-derived
ETF parameters and heat capacity was found for all measured channels.

\begin{table}[htp]
\caption{\label{tab:ETF_parameters}
Summary of calculated ETF parameters.
}
\begin{ruledtabular}
\begin{tabular}{cccccl}
\textrm{Channel}&
$\alpha$&
$\beta$&
$C\,\rm{(fJ/K)}$&
$k_s$&
Notes\\
\colrule
 &  &  &  &  & \\
4 & 66 &1.3& 220 & 0.77 & \textrm{Filter}\\
5 & 77  & 2.9 & 200 & 0.74 & \textrm{Filter}\\
6 & 114  & 4.4 & 200 & 0.80 & \textrm{Filter}\\
12 & 128  & 1.3 & 180 & 0.88 & \textrm{Dark}\\
 &   &  &  &  & \\
11 & 114  & 1.9 & 390 & 0.66 & \textrm{Residual}\\
 &  &  &  &  &\textrm{Power} \\
\end{tabular}
\end{ruledtabular}
\end{table}
Derived values for $\alpha$, $\beta$, the heat capacity
and $k_s$ for Chip 1 used in the modelling  of Figs.~\ref{fig:Response_p11_400} and
\ref{fig:Response_p12_300}
 are given in Table~\ref{tab:ETF_parameters}.
We see that $\alpha$ is low compared to our MoAu TESs, $\beta\sim0.03\alpha$,
and
the responsivity is reduced from the high-$\alpha$ value $k_s=1$.

\section{Power calibration}
\label{sect:Power_calibration}

\subsection{Available in-band power}
\label{sect:Available_power}
We assume that the matched termination load, waveguide and probe behave as an ideal single-mode blackbody
source at temperature $T_\textrm{bb}$ with a bandwidth determined by the waveguide cut-on and probe 3-dB cut-offs giving
a top-hat filter with minimum and maximum transmission frequencies $\nu_{\min}=40$, $\nu_{\max}=65\,\rm{GHz}$ respectively.
Fitting a model to the manufacturer's data we find that loss  in the coax 
can be described by $\alpha_\textrm{coax}(l)\,[\textrm{dB}]=0.2648\, l\,[\textrm{mm}]/\sqrt{\nu\,[\textrm{GHz}]} $,
where $l$ is the
total line length.
Additional losses arise from the 3-dB attenuators
that form the coax heat-sinks (2 being used), and the $1.85\,\,\rm{mm}$ connectors
 (8 in total), each contributing
an additional loss of $\alpha_\textrm{c}=0.223\,\,\rm{dB}$. The total attenuation $\alpha_{l}(\nu)$ (in dB) is then
\begin{equation}
\alpha_{l}(\nu)= \alpha_\textrm{coax}(l) + 6 + 8\alpha_\textrm{c} .
\label{equn:attenuation}
\end{equation}
 The change in available power \textit{at the probe} is
  $\Delta P_{\max}(T_{bb}) =(P_0(T_\textrm{bb})-P_0(T_{0})) 10^{-\alpha_{l}(\nu)/10}$,
  with $T_{0}$ the lowest operating temperature for the blackbody and
\begin{equation}
P_0(T) =\int_{\nu_{\min}}^{\nu_{\max}} h\nu n(\nu,T)    d\nu,
\label{equn:Blackbody_power}
\end{equation}
where $n(\nu,T)= [\exp ( h\nu/(k_\textrm{b} T) -1 ) ]^{-1}$ is the Bose distribution,
 and $h$ is Planck's constant.

To calculate the power transmission to the residual power detector,
we assume each of the filter channels of centre frequency $\nu_{0,i}$, (where $i$ identifies the channel
and $ i \in 1\dots 10$)
 taps its maximum fraction
of the incident  power given by Eq.~\ref{eqn:filter_power_gain} and the transmitted power
is given by Eq.~\ref{eqn:filter_power_transmitted},
with $6\textrm{-dB}$ attenuation at the residual power detector
 due to each filter at its resonant frequency. The  total
available power at the residual power detector $P_r(T_\textrm{bb})$ is
\begin{eqnarray}
&&P_{r} (T)= \eta \int_{\nu_{\min}}^{\nu_{\max}} h\nu n(\nu,T)\,  \prod_i T_i(\nu,\nu_{0,i})\, 10^{-\alpha_{l}(\nu)/10}  d\nu,\nonumber\\
&&
\label{equn:Available_power}
\end{eqnarray}
$\Pi$ is the product and
 $\eta$ is the overall efficiency referenced to the input to the probe.

\begin{figure}[htp]
\begin{center}
\begin{tabular}{c}
\includegraphics[width=8.0cm]{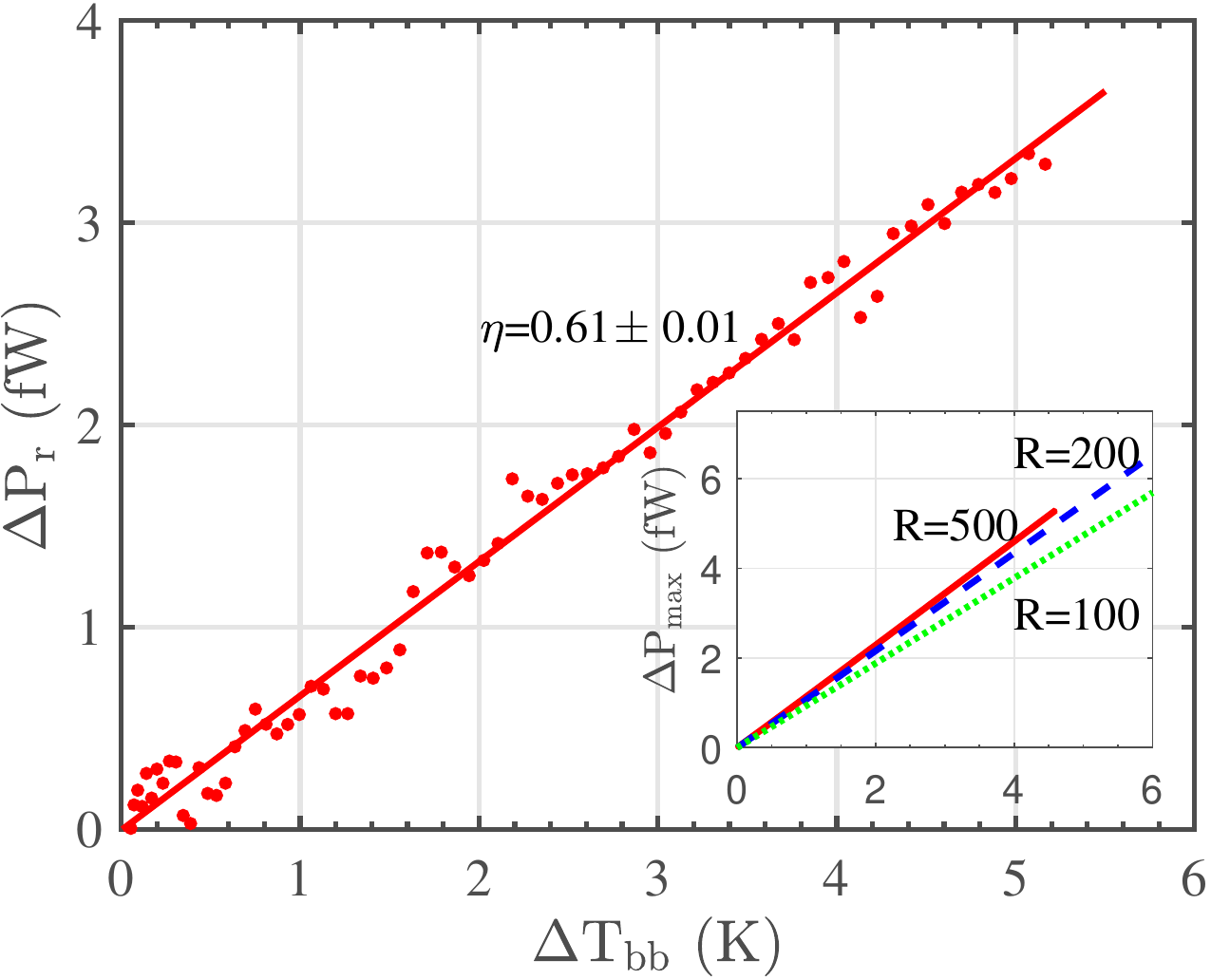}
\end{tabular}
\end{center}
\caption
{ \label{fig:BB_P11}
Power detected for the residual power detector as a function of change in blackbody temperature. The inset
shows the maximum available power $P_{\max}$ for 3 values of the filter-bank resolution.
$\mathcal{R}=500$ is the designed resolution.
 The indicated efficiency  $\eta$
 assumes $\mathcal{R}=200$ and was calculated with the maximum  TES responsivity $s_i=-1/I_0(R_0-R_{\textrm{load}})$.}
\end{figure}

\subsection{Coupled power measurement}
\label{sect:Measured_power}
Measurements of the
detection efficiency were made on Chip 1 with the cold stage temperature maintained at $90\,\,\rm{mK}$ using a PID
feedback loop to regulate the ADR base temperature. Under typical (unloaded) conditions the
cold stage temperature is constant at
$T_\textrm{b}=\pm 100\,\,\mu\textrm{K}$.
$T_\textrm{bb}$ was increased by stepping the blackbody heater current at $~2\,\textrm{Hz}$
with the change in measured TES current and $T_\textrm{bb}$ digitized at $2\,\textrm{kHz}$.
Changes in detected power in both the residual power and dark detectors as $T_\textrm{bb}$ was increased
were calculated from the change in their respective measured currents   assuming the
 simple power-to-current responsivity, $s_I(0)=-1/(I_0(R_0-R_\textrm{L}))$,
  with the TES operating points calculated from measured and known electrical circuit parameters.
At the highest $T_\textrm{bb}$ used, $\sim 10\,\,\rm{K}$,  $T_\textrm{b}$ increased by $\sim 1\,\,\rm{mK}$ suggesting power
loading onto the cold stage.
Under the same blackbody loading, and even after subtracting the
expected dark detector power response due to $T_\textrm{b}$,  $\Delta P_\textrm{dark}=G_\textrm{b,dark}\Delta T_\textrm{b}$,
the dark detector indicated a current (hence power) response that we interpret here as additional incident power.
We modelled this residual response  as a change in the temperature of the Si chip itself $T_\textrm{chip}$, finding an increase
 of order $3\,\,\rm{mK}$ at the highest
$T_\textrm{bb}$.
The modelled chip response $\Delta T_\textrm{chip}$ closely
 followed a quadratic response with $\Delta T_\textrm{bb}$. Although the origins of this
 additional power loading could not be determined in this work,
 it may have arisen from residual short wavelength  loading  from the blackbody source.
  At 10~K the peak in the multimode blackbody
 spectrum is around $1\,\,\rm{THz}$ implying that there may be a significant available detectable power at high frequencies.
 Even a very small fraction of  this power would be sufficient to
 produce the inferred chip heating.
Changes in detected power for the residual power detector with $T_\textrm{bb}$ were accordingly  reduced,
to include
 the effect of both the measured $\Delta T_\textrm{b}$, and the modelled $\Delta T_\textrm{chip}$.

Figure \ref{fig:BB_P11} shows the change in detected power $\Delta P_r$
 as a function of  $\Delta T_\textrm{bb}$, for $T_\textrm{bb}$ in the range
4.2 to $9.4\,\,\rm{K}$. The power response is  close to linear with temperature (the correlation coefficient
$R_{\textrm{cor}}=0.98$). The inset shows the
 maximum available power $P_{\max}$ calculated for three values of the filter resolution $\mathcal{R}_i=100$, $200$
 and $500$ - the designed value.
 The residual power increases as $\mathcal{R}_i$ increases.
 As discussed in Sec.~\ref{sect:Filter_response}
 the designed value may represent an over-estimate of the
 achieved $\mathcal{R}$ for the measured chip.
Assuming $\mathcal{R}=200$ we estimate an efficiency of $0.61\substack{+0.01 \\ -0.03}$ and the lower
error is calculated assuming the designed filter resolution $\mathcal{R}=500$. The measured power
and hence efficiency estimate were calculated assuming the
maximum power-to-current responsivity $s_i=-1/(I_0(R_0-R_\textrm{L}))$.
 From the measured response of the residual power detector described in
Sec.~\ref{sect:ETF_estimates}, and Table~\ref{tab:ETF_parameters}
 we see that this \textit{over-estimates} the sensitivity
  by a factor $1/k_s\simeq 1.5$.
Taking account of this correction, our final estimate of overall efficiency
referred to the input to
the probe is increased to $\eta=0.91\substack{+0.015 \\ -0.05\,}$.

\section{Filter spectral response}
\label{sect:Filter_response}
\begin{figure}[htp]
\begin{center}
\begin{tabular}{c}
\includegraphics[width=8.0cm]{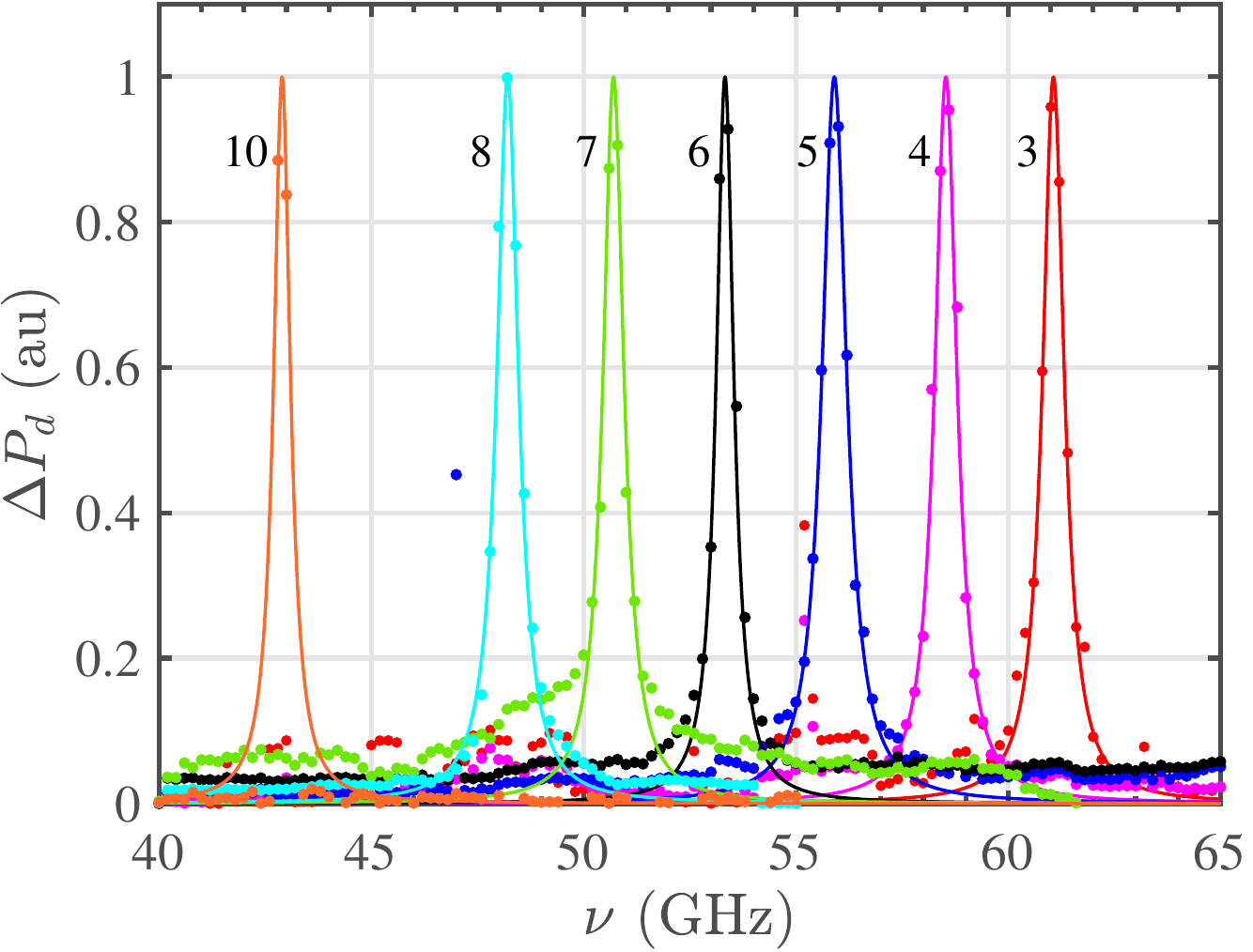}
\end{tabular}
\end{center}
\caption
{ \label{fig:Filter_response}
Normalised power detected for filter channels 3 to 8 and  10 for Chip 2 (dots) and the calculated
filter profiles (full lines). Colours and annotation identify the individual channels.
}
\end{figure}
Measurements of the
spectral response were made on Chip 2.
Figure~\ref{fig:Filter_response} shows (dots) the normalised power detected
and (solid lines) the modelled response
using Eq.~\ref{eqn:filter_power_gain} (i.e. a Lorentzian)
for filter channels 3 to 8 and  10
as labelled. Colours and annotation identify the individual channels.
The individual filter responses $\mathcal{R}_i$ and $\nu_{0,i}$ are shown in  Table~\ref{tab:Filter_parameters}
along with the nominal design filter centre frequencies.
Errors in $\nu_{0,i}$  were calculated from the fit to the data.
The frequency sampling of channel 10 was insufficient to
determine the filter response. For this channel the model fit shown was calculated assuming
$\mathcal{R}_{10}=90.2$ (the mean of the other channels) and is plotted only as a guide to the eye.
For this series of measurements, within measurement noise, there was \textit{no} observable response  in the  dark channel
and we estimate $\Delta P_{\textrm{dark}}<100\,\,\textrm{aW}$.
The absence of any detectable dark response
using a \textit{power-levelled, narrow-band,}
source operating at a fixed temperature of $\sim 300\,\,\textrm{K}$ -- in which case we might
expect no, or only very small changes in additional loading --
gives weight to our interpretation and analysis that the  blackbody and efficiency measurements
should indeed be corrected for  heating from a
\textit{broad-band, temperature modulated} source. In those measurements
 the dark detector showed an unambiguous response.

\begin{figure}[htp]
\begin{center}
\begin{tabular}{c}
\includegraphics[width=8.0cm]{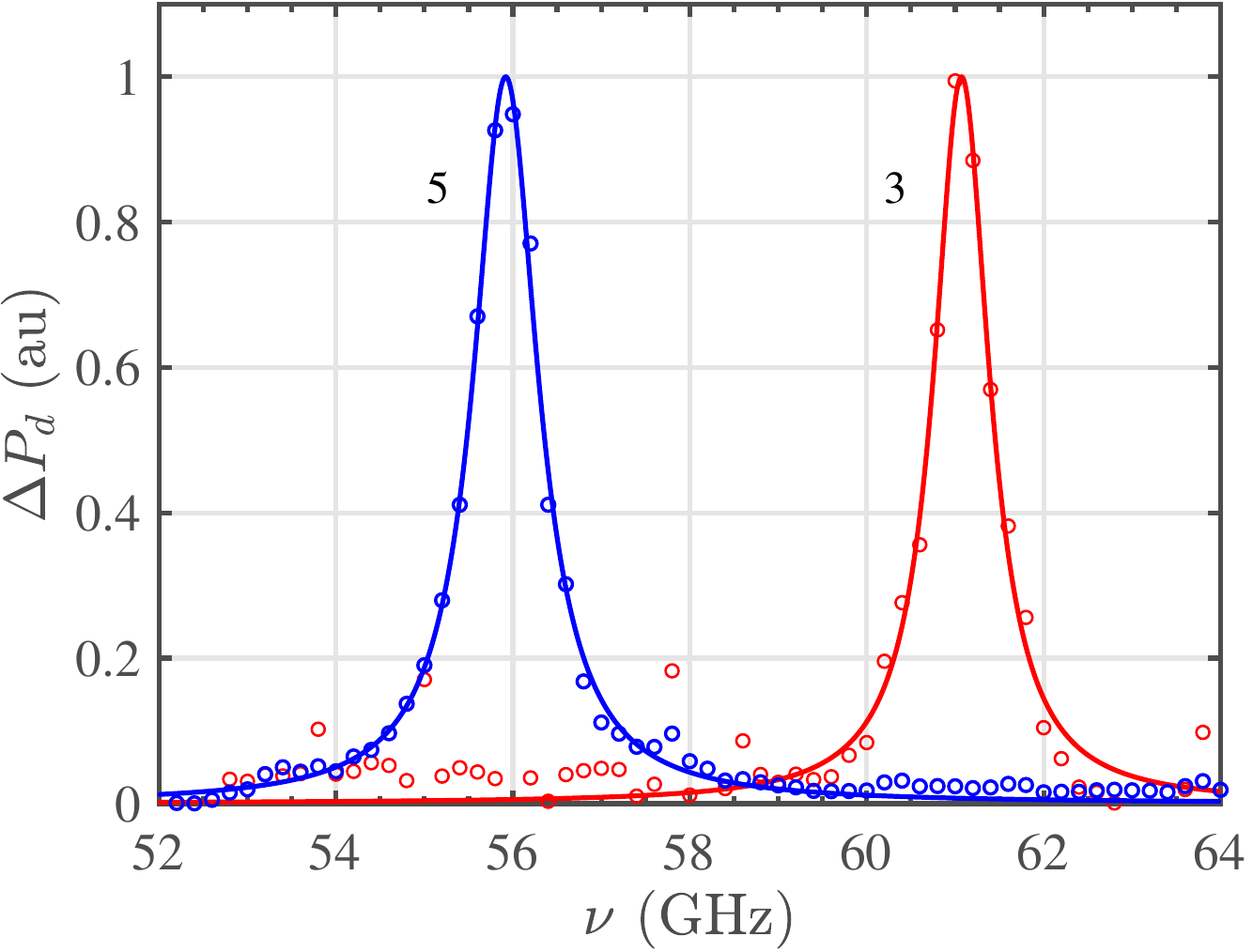}
\end{tabular}
\end{center}
\caption
{ \label{fig:Filter_response_detail}
Power detected from filter channels 3 and 5 with Lorentzian fits. }
\end{figure}

\begin{table}[htp]
\caption{\label{tab:Filter_parameters} Summary of designed and measured filter channel characteristics.
}
\begin{ruledtabular}
\begin{tabular}{cccc}
\textrm{Channel}& $\nu_\textrm{design} $& $\nu_{0,i}$ & $\mathcal{R}_i$\\
 & $\rm{(GHz)} $& $\rm{(GHz)} $ & \\
\colrule
&&&\\
 3 & 60.0 & $61.06\pm0.013$ & $99.0\pm6.5$   \\
 4 & 57.5 & $58.53\pm0.008$ & $92.8\pm3.4$\\
 5 & 55.0 & $55.90\pm0.013$ & $79.3\pm4.2$\\
 6 & 52.2 & $53.33\pm0.003$ & $106.5\pm2.0{\rm{\,\,\,}}$\\
 7 & 50.0 & $50.71\pm0.010$ & $85.3\pm5.5$\\
 8 & 47.5 & $48.21\pm0.005$ & $78.4\pm1.9$\\
 10 & 42.5 & $42.89\pm0.001$ &  \\
\end{tabular}
\end{ruledtabular}
\end{table}
The shapes of the individual filter responses were close to the expected Lorentzian as shown in more detail
in Fig~\ref{fig:Filter_response_detail}
 for channels 3 and 5.

\begin{figure}[htp]
\begin{center}
\begin{tabular}{c}
\includegraphics[width=8.0cm]{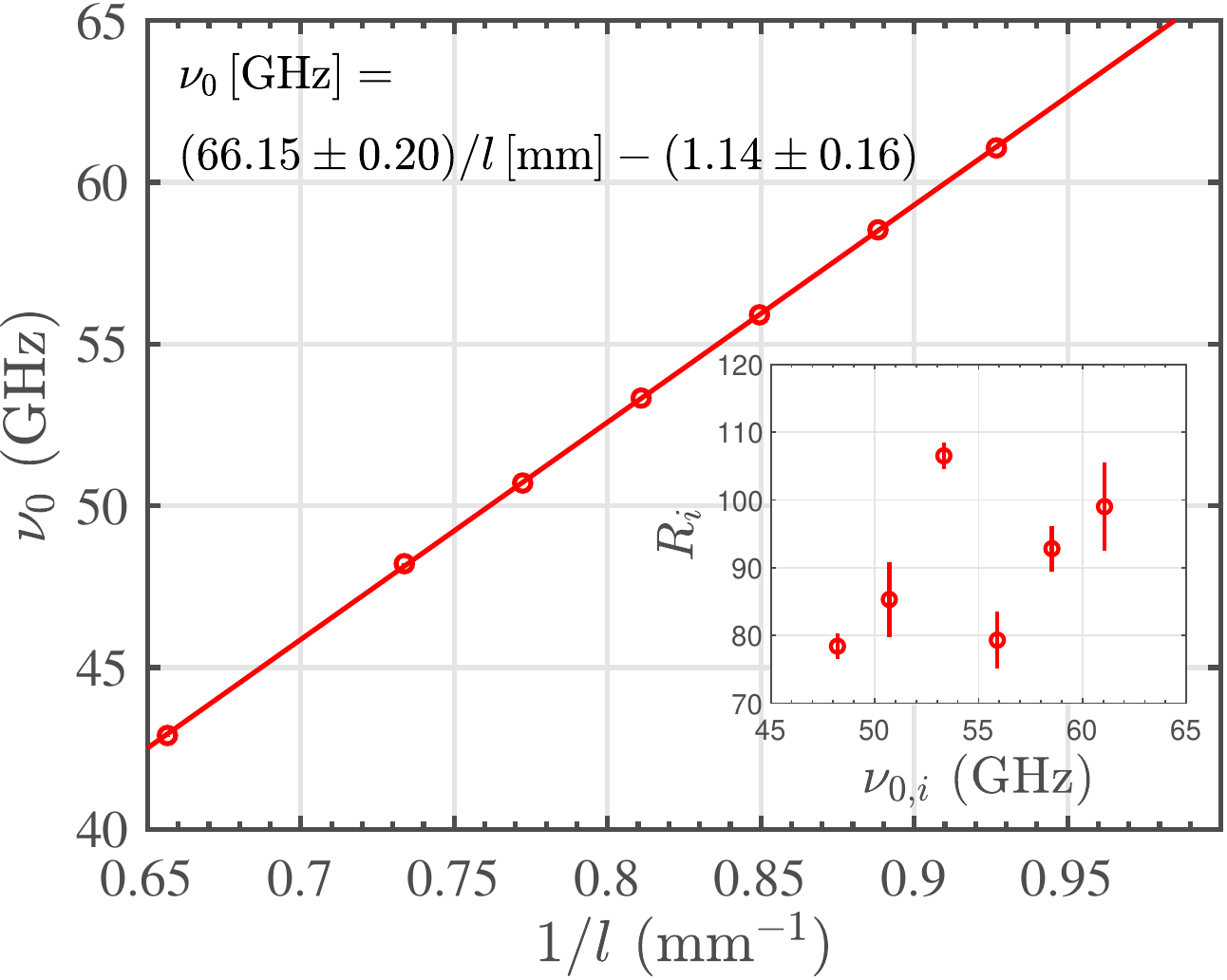}
\end{tabular}
\end{center}
\caption{
 \label{fig:Filter_frequencies}
 Plot of measured filter centre frequency as a function of inverse filter length. (Inset) The measured channel
resolutions $\mathcal{R}_i$ with frequency. }
\end{figure}

Figure~\ref{fig:Filter_frequencies} shows the measured filter centre frequencies $\nu_{0}$ as a function of
reciprocal filter length
$l$. The correlation is close to unity ($R_{\textrm{cor}}=0.9998$)
as predicted by Eqs.~\ref{eqn:resonant_frequency} and \ref{eqn:detuning},
 demonstrating  the precision with which it is possible to
 position the
individual filter channels.
From the calculated regression line
$\nu_{0}\,\textrm{[GHz]}= (66.15\pm0.20)/l\,\textrm{[mm]}-(1.14\pm016)$,
and assuming a modest lithographic precision of
$\pm 1 \,\, \mu{\textrm{m}}$ for the fabrication process,
we calculate $\delta \nu=\pm37\,\,{\textrm{MHz}}$
for a 50~GHz filter, or  $\delta \nu/\nu=0.0007$.
The finite value for the intercept is unexpected
and is an area for future investigation.
Uncertainties in parameters such as the realised coupling capacitance and the exact permittivity
of the dielectric constant are expected, but from
Eqs.~[1] and [4],
should only affect the constant of proportionality in the relationship
$\nu_0 \propto l^{-1}$, rather than generate an offset.
Instead, the offset suggests an issue with some detail  of the underlying circuit model.

\section{Summary and Conclusions}
\label{sect:Conclusions}
We have described the design, fabrication and characterization of a superconducting filter bank spectrometer
with transition edge sensor
 readout. We have described the design of a waveguide-microstrip radial probe transition with wide bandwidth.
We have demonstrated detection of millimetre-wave
radiation with frequencies from 41 to 64~GHz with characteristics that are
already
suitable for atmospheric temperature
sounding using the O$_2$ absorption band. The temperature of a cryogenic blackbody was modulated to determine the
power detection efficiency of these prototype devices, finding an efficiency  $\eta=0.91\substack{+0.015 \\ -0.05}$
using the measured impedance of the
TES to determine the detection responsivity.
Filter profiles were determined on a separate device.
The measured channel profiles were Lorenztian as expected. We measured
extremely good predictability of relative channel
placement, and  also in individual channel placement
$\delta \nu/\nu=0.07\,\%$.
We found somewhat lower filter resolutions $\mathcal{R}$ than the designed values, possibly arising from our use
in the
filter design
of
dielectric constant measurements at lower frequencies compared to those explored here
 or, perhaps, unexpected dielectric losses.
 This will be an area for future investigation.
In the next phase of this work we will continue to investigate this question but
also investigate alternative filter architectures -- particularly coplanar structures
where  dielectric losses may-well be lower.
Guided by the measurements reported, we will
refine our designs to increase channel density for the O$_2$-band observations and
 also include higher frequency channels
 particularly aimed at observation of
 the 183~GHz water-vapor line to enable a complete on-chip atmospheric temperature
 and humidity sounding instrument.
Finally we  emphasize that, although we have presented and discussed these measurements in the context
of an enhanced atmospheric temperature sounding demonstrator,
we believe we have already shown impressive spectroscopic and  low-noise  power detection
 at technically challenging
 millimetre-wave frequencies, with applications across a broad spectrum
of scientific enquiry.
\section{Acknowledgements}
This work was funded by the UK CEOI under Grant number RP61060435A07.
\bibliography{TESreferences7}

\end{document}